\documentclass[prd, twocolumn, nofootinbib, floatfix]{revtex4-1}

\usepackage{amsmath}
\usepackage{graphicx}
\usepackage{dcolumn}
\usepackage{bm}
\usepackage{epsfig}
\usepackage{amssymb,latexsym,mathrsfs}
\usepackage{graphicx}
\usepackage{color}
\usepackage{hyperref}

\usepackage{tabu}

\hypersetup{
    colorlinks=true,
    linkcolor=red,
    citecolor=blue,
} 

\newcommand{\be}{\begin{equation}}
\newcommand{\ee}{\end{equation}}
\newcommand{\bs}{\begin{split}} 
\newcommand{\bea}{\begin{eqnarray}}
\newcommand{\eea}{\end{eqnarray}}

\newcommand{\neff}{N_{\rm eff}}

\newcommand{\sumu}{\Sigma m_\nu} 
\newcommand{\mnu}{\Sigma m_\nu} 
\newcommand{\mpci}{\,{\rm Mpc}^{-1}}

\begin{document}

\title{Inflationary Freedom and Cosmological Neutrino Constraints} 
\author{Roland de Putter$^{1,2}$, Eric V.\ Linder$^{3,4}$, Abhilash Mishra$^2$} 
\affiliation{$^1$Jet Propulsion Laboratory, California Institute of Technology, Pasadena, CA 91109\\
$^2$California Institute of Technology, Pasadena, CA 91125\\ 
$^3$Berkeley Lab \& Berkeley Center for Cosmological Physics, 
University of California, Berkeley, CA 94720, USA\\ 
$^4$Institute for the Early Universe WCU, Ewha Womans 
University, Seoul 120-750, Korea}

\begin{abstract}
The most stringent bounds on the absolute neutrino mass scale come from
cosmological data.
These bounds are made possible 
because massive relic neutrinos
affect the expansion history of the universe
and lead to a suppression of matter clustering on
scales smaller than the associated free streaming length.
However,
the resulting effect on cosmological perturbations
is relative to the primordial power spectrum of density perturbations from 
inflation, so freedom in the primordial power spectrum affects neutrino mass 
constraints. 
Using measurements of the cosmic microwave background, the galaxy power spectrum
and the Hubble constant, we constrain neutrino mass and number of species for 
a model independent primordial power spectrum.
Describing the primordial power spectrum by a 20-node spline,
we find that the neutrino mass upper limit is a factor three weaker than when a power law form is
imposed, if only CMB data are used.
The primordial power spectrum itself is constrained to better than $10 \%$ in the wave vector range
$k \approx 0.01 - 0.25$ Mpc$^{-1}$. Galaxy clustering data
and a determination of the Hubble constant play a key role in 
reining in the effects of inflationary freedom on neutrino constraints. The inclusion of both
eliminates the inflationary freedom degradation of the neutrino mass bound, giving for the sum of neutrino masses
$\mnu < 0.18$ eV (at $95 \%$ confidence level, Planck+BOSS+$H_0$),
approximately independent of the assumed primordial power spectrum model.
When allowing for a free effective number of species, $\neff$, the joint constraints on $\mnu$ and $\neff$
are loosened by a factor 1.7 when the power law form of the primordial power spectrum
is abandoned in favor of the spline parametrization.

\end{abstract}

\date{\today} 

\maketitle

\section{Introduction} 

Cosmology has revealed rich structure beyond the Standard Model of particle 
physics, with dark matter, an inflationary acceleration of expansion in the early 
universe and a dark energy acceleration in the late universe. Moreover, 
we now know that a required 
extension to the standard models of both particle physics and cosmology 
is the presence of neutrino mass.  Laboratory 
neutrino oscillation experiments indicate that at least two species must 
be massive, and the sum of all three species must be $\sumu>0.055\,$eV 
\cite{pdg}.  

The strongest upper bounds arise from cosmological 
measurements sensitive to the suppression of matter density perturbations
caused by neutrino free streaming and to the effect of neutrino mass on 
the expansion history of the universe.  
These bounds, however, start 
from assumptions about early universe physics, such as a power law form 
for the primordial power spectrum (PPS).  Restricting the form in this way 
will tighten the neutrino limits and, if there exists deviation from a power 
law, bias the results. 
While a power law PPS is predicted by the simplest models of inflation, 
there exist other models of inflation with non-trivial features in the 
PPS. As such, it is important to investigate the neutrino bounds 
without making assumptions about (unknown) inflationary physics. 

The key cosmological observable (in linear perturbation theory) is
the late time power spectrum $P(k)$ (or angular power spectrum $C_\ell$), which is a convolution of the 
PPS $\Delta^2_{\mathcal R}(k)$ that encodes information about inflationary 
physics and the transfer function of cosmological perturbations. 
The transfer function can be
calculated from well tested physics (linearized gravity and photon-baryon 
fluid physics)
and is described by a small number of cosmological parameters,
including neutrino mass and effective number of species,
specifying the energy budget and ionization history of the universe.
The PPS on the other hand depends on physics at an energy scale
that has never been directly tested. Usually, it is parametrized as a scale invariant
power law inspired by the simplest models of inflation. 
However, assuming a particular functional form for the PPS can 
bias the estimates of cosmological parameters 
(e.g.\ \cite{kinney01}\cite{hazraetal13}). 

In this paper 
we focus on how the constraints on neutrino mass and effective number of 
species are affected when more freedom is allowed in the form of the PPS. 
Indeed, even enlarging the parameter space from a simple power law PPS 
to one with running (scale dependence) of the slope 
can strongly affect the neutrino bounds.  For example, 
\cite{zhaoetal13,1206.0109} found significant covariance between running 
and both neutrino mass and effective number of species. 

Due to the scale dependent effect of neutrino mass on the perturbation power 
spectrum, properties of neutrinos and of the inflaton responsible for 
generating the PPS enter in the data tied closely together in ways different 
from other extensions to the Standard Model like dark matter and dark energy. 
Many inflation theories in the post-Planck data era do have spectral features 
deviating from a simple power law. These may include oscillations, steps, 
and bumps and can arise from physics such as axion monodromy 
\cite{eastflaug13}, 
holography \cite{eastetal11}, sound speed \cite{achuetal13}, among many other 
current ideas. 

Several approaches exist in the literature for reconstructing a free PPS. 
These include the ``cosmic inversion'' method 
\cite{matsumiya02, matsumiya03, kogo05, nagata08, kogo04}, 
regularisation methods like truncated singular value decomposition 
\cite{nicholson10} and 
Richardson-Lucy iteration \cite{hazraetal13,shafieloo04, Nicholson09}, and maximum 
entropy deconvolution \cite{goswami13}. 
Recently \cite{hunt13} carried out a reconstruction of the PPS 
employing Tikhonov regularisation using multiple datasets and detected 
several features in the PPS at a 2$\sigma$ level of significance (also 
compare \cite{hazraetal13}).

Here we focus on the issue of the dependence of neutrino constraints on 
assumptions about the PPS, rather than reconstructing the PPS per se.
Suppression of growth due to massive neutrinos enters around the free streaming scale, or
comoving
wavenumbers $k\gtrsim0.01\,{\rm Mpc}^{-1}$.
Moreover, power spectra of cosmic perturbations are fundamentally observed as a function of angular scales
(and redshifts, in case of three dimensional large scale structure),
so that the effect of massive neutrinos on
the expansion history can in principle cause changes in the observed CMB and large
scale structure power spectra on all scales by shifting these spectra horizontally.
Variations in primordial power on all observable scales,
from approximately
the cosmic horizon scale, down to small scales where Silk damping (CMB) or
non-linear clustering (large scale structure) degrades the cosmological information,
may therefore affect neutrino constraints.
Thus we investigate how freedom in the PPS over some 2.5 orders 
of magnitude, $k\approx 0.001-0.35\,{\rm Mpc}^{-1}$, affects determination 
of the neutrino and cosmological parameters.

For our purposes of investigating effects on the cosmological parameters 
we want a robust, model independent description. Examples include 
the use of wavelets \cite{mukherjee103,mukherjee203,mukherjee303,mukherjee05}, 
principal components \cite{leach06}, tophat bins with no interpolation 
\cite{wang99}, linear interpolation 
\cite{wang02,bridle03,hannestad03,bridges05,spergel07,bridges06,bridges09}, 
smoothing splines \cite{seaflon05, verde08, peiris10, gauthier12}, 
and power-law bins \cite{hannestad01}. 
We choose to describe the PPS by cubic splines, following 
\cite{hlozeketal12}. This preserves 
model independence while encompassing the power law model, and is smooth.  

A completely free form PPS could exactly mimic (at least within 
cosmic variance) neutrino mass effects 
for a single type of observations, e.g.\ cosmic microwave background (CMB) 
temperature perturbations, though possibly requiring order unity sculpting 
of the PPS \cite{kinney01}. Since other observations, such as CMB polarization 
spectra or matter density power spectra, enter with different redshift-weighted transfer functions
relative to the PPS, and since these transfer functions depend in different ways on
cosmological parameters,
then combining power spectra data (or external constraints on other 
cosmological parameters) plays an important role in fitting both inflaton 
and neutrino properties (see \cite{TegZal02}). 

We therefore carry out several studies on how freedom in the PPS affects 
neutrino constraints: we consider CMB data alone, with inclusion of large 
scale 
structure data, and with inclusion of Hubble constant measurement. As well we 
investigate the standard scenario with three neutrinos with unknown total 
mass, and also the scenario with a free total mass {\it and} a free effective 
number of neutrino species, $\neff$. 

Section~\ref{sec:model} 
discusses the treatment of the PPS in a substantially model independent 
manner. 
We describe the data used and our method for calculating parameter
constraints in Sec.~\ref{sec:datamethod}. Cosmological 
parameter estimation results are presented in Sec.~\ref{sec:par}, examining 
the covariance between standard, neutrino, and extended PPS parameters, and 
the role of CMB, large scale structure, and Hubble constant data.
We summarize our results in Section \ref{sec:concl} and discuss the sensitivity of our
constraints to the assumed PPS spline parametrization and to the CMB data included in the Appendices.

\section{Primordial Power Spectrum} \label{sec:model}

In the inflationary scenario for generation of density perturbations, 
the universe is in a near-de Sitter state where the quantum fluctuations 
of the inflaton field produce scalar (gravitational potential) and tensor 
(gravitational wave) metric perturbations.  In simple, single field models 
of inflation the gravitational potential perturbations are Gaussian and 
nearly scale invariant.  This implies that they (and the density 
perturbations through the Poisson equation) are completely characterized 
by the two-point function or power spectrum.  Since inflation must end, 
the spacetime is not exactly de Sitter and so the power spectrum is not 
exactly scale invariant. 

For slow rolling of the inflaton field value over time, the 
PPS is conventionally expanded in a Taylor series about the value at 
some pivot scale, 
\be 
\Delta^2_{\mathcal R}(k)=\Delta^2_{\mathcal R}(k_0) 
\left(\frac{k}{k_0}\right)^{n_s-1+(\alpha_s/2)\ln (k/k_0)} \ , 
\ee 
where $\Delta^2_{\mathcal R}$ is the curvature perturbation power spectrum, 
$k_0$ the pivot wavenumber, $n_s$ the tilt, and $\alpha_s$ the running. 
However, in more general inflation scenarios the slow, smooth evolution 
of the field can be replaced with faster variations, oscillations, and 
features (which can in particular circumstances be treated through a 
generalized slow roll formalism \cite{dvorkinhu10}). 

Forcing a power law form could bias the results for all quantities, even 
the late time cosmological parameters, and certainly affects the uncertainty 
of their estimation. Given that current cosmological data are severely 
constraining the sum of neutrino masses from above, with this bound beginning 
to approach the lower limit imposed by neutrino oscillation terrestrial 
experiments, it is worthwhile exploring the link between inflationary 
assumptions and neutrino constraints. 

Therefore we attempt a model independent approach where no functional 
form is assumed. The values of the PPS amplitude at various wavenumbers 
(nodes) are allowed to float freely, and these are smoothly connected 
using a cubic spline.
With enough nodes this can give an excellent 
approximation to a wide range of functions, including nonmonotonic and 
oscillatory behavior. 
The typical number of nodes used in the literature is around 20 though it 
can be as high as 50 \cite{Ichiki09}; we use 20, although we explore the 
effects of using 10 or 40 in Appendix \ref{sec:nodes}. The number of nodes (parameters) is thus much 
less than the number of data points and an MCMC analysis is expected to 
give accurate confidence limits on cosmological parameters. 

We first define a normalized primordial power spectrum $p(k)$,
such that 
\be
\Delta^2_{\mathcal{R}}(k) \equiv \Delta^2_{\mathcal{R},0} \, p(k)= 
\Delta^2_{\mathcal{R},0} \times {\rm spline}[p\{k_i\}] \ , 
\ee 
where we choose $\Delta^2_{\mathcal{R},0} \equiv 2.36 \times 10^{-9}$ (the approximate value of the primordial power spectrum
amplitude preferred by current data), 
such that $p(k)$ is expected to be of order unity. Note that the actual 
amplitude can vary without loss of generality by changing $p(k)$. 
We then specify the values $p_i \equiv p(k_i)$ of this normalized PPS
at a set of $N$ spline nodes, $k_i$.
At $k < k_1$, we fix $p(k)= 1$, whereas at $k > k_N$,
we set $p(k) = p_N$. In the intermediate range,
$p(k)$ is given by a cubic spline. 

To encompass the range of scales well probed by CMB and galaxy clustering 
data we take $k_1=0.001\mpci$ and $k_N=0.35\mpci$. 
The low end is slightly larger than the wave vector
corresponding to the cosmic horizon, and is thus close to the smallest
$k$ that could be probed by {\it any} observable. 
For the spacing of the $k_i$ nodes we follow 
\cite{hlozeketal12}, using $N=20$ nodes with a logarithmic spacing, such that 
$k_{i+1} = 1.36 \, k_i$. 
This allows the PPS to cause variations in the CMB and galaxy power spectra 
on scales comparable to those associated with the features (such as baryon 
acoustic oscillations) introduced by the transfer functions depending on 
cosmological parameters, and hence we can explore how PPS freedom interacts 
with cosmological parameter estimation. 
In Appendix \ref{sec:nodes} we will consider alternative choices of the PPS 
characteristics to test the robustness of the results. 

We allow the PPS parameters $\{ p_i \}$ to vary in the range $0.01 < p_i < 10$
(imposing uniform priors). When the node parameters are close to zero, it is possible
for the spline to return negative values for $p(k)$ at some intermediate $k$. To avoid this unphysical behavior
of the PPS, we restrict to $p(k) \geq 0.01$, setting $p(k) = 0.01$ whenever the spline returns a value smaller than $0.01$.
Our results are insensitive to the exact choice for the lower bound of the $p_i$ range and to
the details of the cutoff because primordial power spectra with nodes $p_i \sim 0.01$
have a very low likelihood.
Given the PPS, the CMB temperature and polarization power spectra and the
matter power spectrum at any redshift are
obtained by convolving the PPS with the transfer functions
for CMB multipoles and matter density perturbations, in the usual manner.

\section{Data and Method} \label{sec:datamethod} 

Our choice of CMB data closely follows that
of the Planck collaboration \cite{planckcosmoparam,plancklike}.
We use the Planck temperature power spectrum, together with high resolution ({\it high-}$\ell$)
temperature data from the Atacama Cosmology Telescope (ACT) \cite{actdasetal13,actdunkleyetal13}
and the South Pole Telescope (SPT) \cite{sptreichardt12,sptkeisler11}.
We also include low multipole ($\ell < 23$) polarization data from WMAP (referred to as WP
in the Planck papers) \cite{wmapbennettetal13}. In the standard case of a power law PPS,
the latter data set mainly serves to constrain the optical depth due to reionization,
$\tau$, which is otherwise strongly degenerate with the amplitude of primordial perturbations, $A_s$.
For a free PPS, the polarization data play a more important role, as discussed in Appendix \ref{sec:cmbpol}.
The temperature power spectrum measurements from Planck, ACT and SPT are illustrated in
Figure \ref{fig:pps cl mnu} (see Section \ref{sec:par} for more discussion of this figure).

We incorporate the required set of 31 ``nuisance'' parameters
needed to take into account foregrounds, beams and calibration uncertainties.
These parameters will be marginalized over when parameter constraints are computed.
The main use of the {\it high}-$\ell$ data is to help constrain a number of these nuisance parameters
describing extragalactic foregrounds,
thus improving constraints on cosmological parameters that are partially degenerate with the ``nuisance'' parameters
when using Planck data only.

In order to keep the number of observables small and the interpretation of our results clean,
we do not include the reconstructed lensing potential power spectrum data 
from Planck.
However, the effect of gravitational lensing on the CMB temperature power spectrum {\it is\/}
modeled and in fact contributes strongly to the constraints on neutrino mass in the case of a power law primordial
power spectrum \cite{planckcosmoparam}.

Physically, neutrino mass acts to suppress power in the photon or matter 
density perturbations, damping small scale power. Moreover,
massive neutrinos affect the expansion history and thus cosmic distances, modifying the projection
from physical scales to observed angles and redshifts. For a free form PPS one 
can imagine a complete degeneracy could arise between a power law primordial 
spectrum with some neutrino mass and an appropriate PPS with a different 
neutrino mass.  A similar confusion could occur for the number of effective 
neutrino species, although this can also affect larger scales through 
changing the time of matter-radiation equality.  However, since these are 
one time adjustments, measurements of the perturbation power spectrum with 
different redshift weightings would become 
distinguishable.  Such different weights occur in the polarization, lensing, 
or matter power spectra relative to the temperature power spectrum. 

One of the goals of this article is to see to what extent combining the high redshift measurement
from the CMB with a low redshift measurement of large scale structure
can constrain neutrino properties (and other cosmological parameters)
without making strong assumptions about the primordial power spectrum.
The current state of the art in large scale structure surveys is the Baryon Oscillation
Spectroscopic Survey (BOSS) \cite{dawsonetal13}.
Therefore, we use the angle-averaged galaxy density power spectrum obtained from the BOSS CMASS
sample (see, e.g.~\cite{whiteetal11,andersonetal12}) made available with data release 9 (DR9, \cite{dr9}).
This sample has an effective redshift $z_{\rm eff} = 0.57$ and covers 
an effective volume $V_{\rm eff} = 2.2$ Gpc$^3$.
The observed power spectrum is shown in Figure \ref{fig:pps pkg} (see Section \ref{sec:par}
for further discussion of this figure).
We include only the black data points in our likelihood, corresponding to the wave vector range
$k = 0.03 - 0.12 h/$Mpc.

In the likelihood analysis, we marginalize over possible systematic contributions
to the large-scale power spectrum, by subtracting from the observed spectrum a template,
$P_{g,{\rm obs}}(k) \to P_{g,{\rm obs}}(k) - S P_{\rm sys}(k)$, where $P_{\rm sys}(k)$
is the template and $S$ a free parameter with prior range $S = [-1,1]$. We refer to
\cite{rossetal12,rossetal13} for details. For each set of cosmological and nuisance parameters,
we then compare the observed spectrum to a theory spectrum.
Since we restrict the analysis to linear and only mildly non-linear scales,
it is appropriate (see e.g.~\cite{swansonetal10,zhaoetal13}) to model this theory power spectrum
using the simple ``P-model'',
\be
P_g(k) = b^2 P_m(k) + P_0.
\ee
Here, $b$ is a free, scale-independent galaxy bias and $P_m(k)$ the linear matter power spectrum.
The additional free parameter $P_0$ is included to describe deviations from scale-independent bias,
which can arise from a combination of non-linear galaxy bias, non-linear redshift space distortions and
stochastic bias.
Finally, this theory spectrum is convolved with the survey window function to
match the expectation value of the spectrum estimated from the data, see
\cite{zhaoetal13} for details.
Our analysis of the BOSS power spectrum is identical to that in \cite{DAO13}.

Finally, we consider the inclusion of a prior on the Hubble constant. 
We use the value $H_0 = 73.8 \pm 2.4$ km/s/Mpc by \cite{hst11} (R11) based on
supernova distances, which are calibrated using Cepheids observed with the Hubble Space Telescope.
We will refer to this measurement as {\it HST\/}. 
Since the CMB is mostly 
sensitive to the physical densities, while the matter clustering is more 
affected by the fractional densities, then the Hubble constant can play an 
important role in linking the two. Moreover, the CMB data alone leave a strong anti-correlation between
neutrino mass and $H_0$ so that the inclusion of a tight $H_0$ prior can be expected to have a significant effect
on the neutrino mass bound.

While the $H_0$ measurement by R11 has been widely used in cosmological analyses,
some caution is in order with regards to constraints based on this measurement.
First of all, there is a significant and well known tension
between the above direct measurement and the value inferred from the Planck data
in the context of the standard $\Lambda$CDM model \cite{planckcosmoparam} (regardless of whether neutrino mass is a free
parameter).
While this might be a sign of new physics, such as the existence of additional relativistic species,
it could also point to a problem with the data (analysis).
Secondly, the R11 Cepheid data have recently been reanalyzed \cite{E13}
(taking into account the revised geometric maser distance to NGC4258 and modifying the treatment of outliers and Cepheid
metallicity),
leading to a non-negligible shift in the Hubble constant, $H_0 = 72.5 \pm 2.5$ km/s/Mpc.
In Section \ref{sec:cmbgalh0}, we will therefore
also present neutrino mass constraints using the revised $H_0$ measurement.

We modify CAMB \cite{camb} to use the set $\{p_i\}$ as a replacement 
for the usual amplitude $A_s$ and tilt $n_s$ (and running $\alpha_s$).
We obtain both the CMB power spectra and the low redshift matter power spectrum (which in turn is used
to compute the galaxy power spectrum) from this modified version
of CAMB.
In addition to the remaining four standard vanilla cosmology parameters of 
the physical baryon density $\omega_b\equiv\Omega_b h^2$, cold dark matter 
density $\omega_c\equiv\Omega_c h^2$, cosmological constant density $\Omega_\Lambda$ (or, equivalently, Hubble parameter $H_0$), and 
reionization optical depth $\tau$, we include parameters
to describe the neutrino sector. We consider two scenarios: one where
the effective number of neutrino species is fixed to the standard model value
$N_{\rm eff} = 3.046$ and only the sum of 
neutrino masses $\sumu$ is a free parameter, and one where
$N_{\rm eff}$ is left free as well.

For the standard scenario with three species, we assume a degenerate hierarchy, with each neutrino
having a mass $\mnu/3$. With $\neff$ free and $\neff > 3.046$, we simply add an effective number of $\neff - 3.046$ {\it massless} neutrino species
to the three massive ones. When $\neff < 3.046$, we lower the temperature of the three standard neutrinos to obtain the energy density corresponding to
$\neff$. In this regime, the parameter $\mnu$ should be interpreted as a rescaled sum of neutrino masses, $(\neff/3.046)^{3/4} \mnu$.

Thus there are either 25 or 26 cosmological parameters overall, plus 34 (31 for the CMB and 3 for the galaxy power spectrum)
``nuisance'' parameters from the data.  We will be particularly interested 
in the covariance between the PPS parameters and the neutrino parameters; 
that is, how much the relaxation of an assumed power law form influences 
the neutrino constraints. Modifying CosmoMC \cite{cosmomc}, we carry out a Markov chain Monte Carlo 
analysis for parameter estimation, sampling the cosmological parameters including
PPS parameters, and the nuisance parameters.

Before moving on to the results, we note that, generally, the constraints on neutrino properties come both from
the suppressed growth of cosmic perturbations {\it and} from the expansion history.
The latter relates to cosmic distances
$d(z) = \int dz/H(z)$ and the Hubble rate $H(z)$, which, through the Friedmann equation, is proportional
to the (square root of the) total energy density of the universe. Since the present neutrino energy density is proportional
to the sum of neutrino masses, $\omega_\nu \equiv \Omega_\nu h^2 \approx \mnu/94$eV,
and since the relativistic neutrino energy density is proportional to $N_{\rm eff}$,
both cosmic distances and the Hubble rate probe neutrino mass.

\section{Cosmological Constraints} \label{sec:par}

\subsection{CMB-only constraints on neutrino mass}

\begin{figure}[htbp!]
\includegraphics[width=\columnwidth]{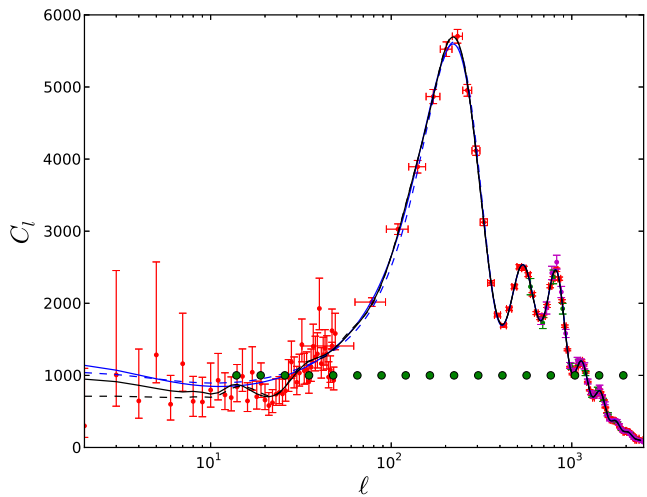}
\caption{The CMB temperature power spectrum is plotted with data points 
from Planck in red, from ACT in green and from SPT in magenta.
The large green dots indicate the locations of the spline nodes. 
The blue (black) solid curve is the best-fit theory spectrum to the 
{\it CMB\/} data set in the case of a power law (free/splined) PPS,
for fixed $\Sigma m_{\nu} = 0$ eV.
The respective dashed curves show the best fit spectra
for fixed $\Sigma m_{\nu} = 2.5$ eV. 
The freedom in the spline PPS can compensate for the neutrino mass 
to keep the power spectrum at $\ell > 10$ virtually the same as the zero 
mass case, showing how inflaton freedom affects neutrino constraints. 
}
\label{fig:pps cl mnu}
\end{figure}

We first consider constraints from the CMB-only combination of data sets. 
The temperature power spectrum measurements are shown in 
Figure \ref{fig:pps cl mnu}. 
For comparison, the solid black (spline PPS) and blue (power law PPS) curves 
show the best-fit theory power spectra for fixed $\sumu = 0$. 
These are in excellent agreement on multipoles $\ell\gtrsim60$, and the 
spline PPS has the freedom to fit variations in the data at smaller 
multipoles. The dashed curves then show the effect of $\sumu=2.5\,$eV. This 
noticeably changes the CMB power spectrum and hence can be ruled out in 
the power law PPS case at high\footnote{A mass $\sumu = 2.5\,$eV is ruled out
at (much) more than $3\sigma$, but the exact significance is difficult to quantify
as the chains do not contain any points beyond $\sumu = 2.5\,$eV.} 
significance. 
However, there can be enough freedom in general in 
the PPS to allow even such a large sum of neutrino masses, as shown by the 
agreement of the dashed and solid black (spline PPS) curves.
This is reflected by the fact that, as we will see, $\sumu = 2.5\,$eV lies within
the $99.7 \%$ CL region for the spline PPS (although outside the $95 \%$ CL region).

The green dots indicate the multipoles corresponding to the projected spline node
wave vectors, $\ell_i = k_i\,D_{LSS}$, where $D_{LSS}$ is the comoving distance to the CMB last scattering
surface. While the mapping between $k$ and $\ell$ is in reality not one-to-one,
i.e.~power at a given wave vector $k$ translates to power at a range of multipoles
$\ell$ rather than just $\ell_i$, the $\ell_i$'s give an impression of where 
a given PPS node modifies 
the angular temperature power spectrum.
Specifically, one sees that the PPS spline is flexible enough to
affect the temperature spectrum across the full range of multipoles constrained by the data
(the signal-to-noise in the multipoles below our lowest node, $\ell \lesssim 10$, is small),
and
to alter the spectrum on the scale of the acoustic oscillations, like 
late time cosmological parameters.

\begin{figure*}[htbp!]
\includegraphics[width=0.95\textwidth]{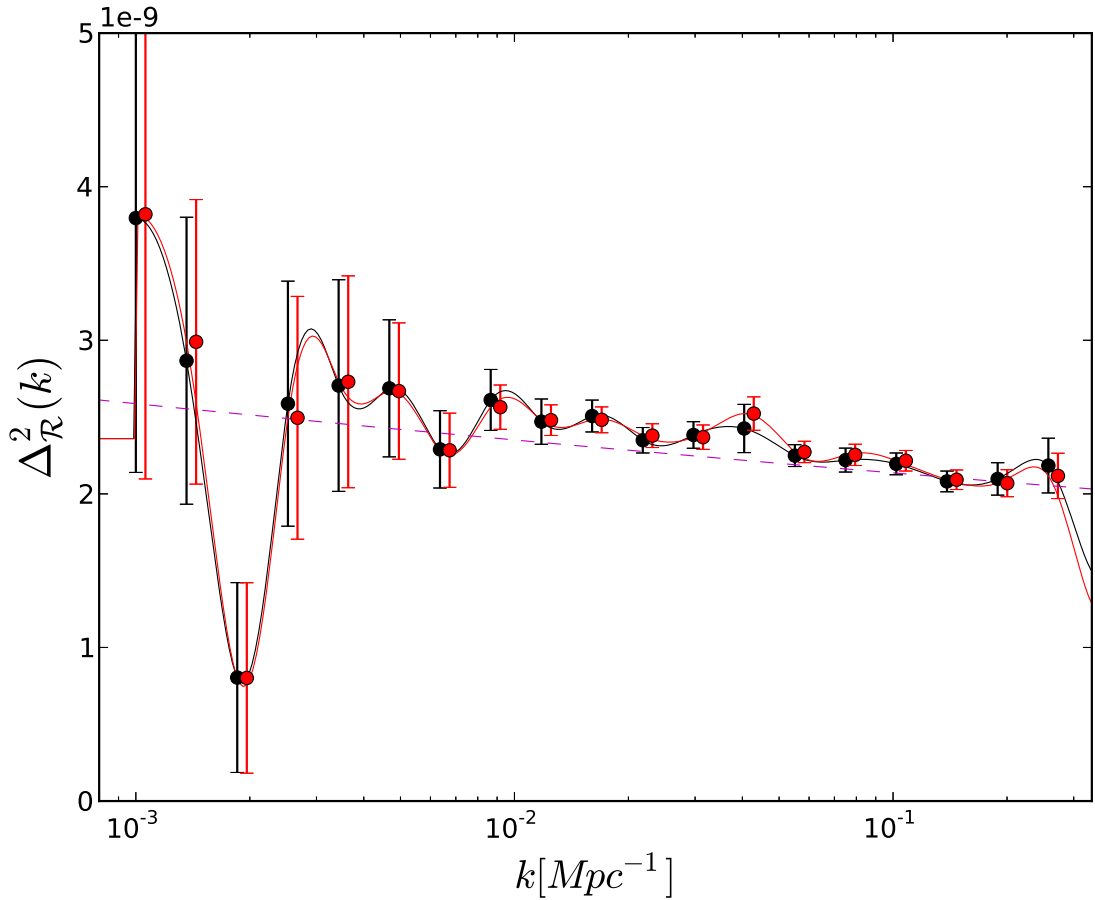} 
\caption{The mean PPS node values are shown, including
error bars. 
The black points and error bars indicate the fit to CMB data only, and the results in light red correspond to CMB + BOSS
galaxy clustering (points slightly offset for clarity).
The solid lines are the primordial power spectra corresponding to the node values shown. 
The dashed line is the best-fit power law spectrum to the {\it CMB\/}-only 
data, marginalized over other parameters.} 
\label{fig:pps constraints p(k)}
\end{figure*}

Figure~\ref{fig:pps constraints p(k)} shows that, despite the large freedom allowed in its shape,
the CMB data place strong constraints on the PPS. The black points and error bars show the mean
PPS values at the spline nodes and the uncertainties.
The black curve
represents the corresponding best-fit PPS at all $k$ (i.e.~the cubic spline going
through the nodes shown in the figure). These PPS constraints are obtained while simultaneously
fitting for the cosmological parameters and the nuisance parameters associated with the CMB data.
The dashed straight line in Figure \ref{fig:pps constraints p(k)} depicts
the power law PPS best fitting
the {\it CMB} data set for comparison.

Except for the highest $k$ node ($k = 0.35$ Mpc$^{-1}$, not shown in 
Figure \ref{fig:pps constraints p(k)}),
all node powers are reasonable well constrained. 
The nodes in the range $k = 0.009 - 0.26$ Mpc$^{-1}$ 
all have $< 10 \%$ uncertainties (relative to the fiducial amplitude $\Delta^2_{\mathcal{R},0} = 2.36 \cdot 10^{-9}$), with the best-constrained node (at $k = 0.14$ Mpc$^{-1}$)
being measured with $3 \%$ precision. The $p_i$ parameters are strongly correlated among themselves, with correlation 
coefficients up to $|\rho| = 0.93$. Moreover, it is not only the nearest neighbors that are strongly correlated 
(or anti-correlated), but the correlations persist for pairs of nodes well separated in $k$ space.

The reconstructed power spectrum displays no strong evidence for deviations from a power law PPS.
The spline PPS model does lead to a better fit to the data, with $\Delta \chi^2 = 33.8$.
Given that the spline PPS has 18 parameters more than the power law model,
this means that the spline model gives a slightly bigger improvement in the fit to the data than
expected purely based on the larger number of free parameters, but the difference
is not significant enough to claim strong evidence in favor of the spline model. 
Some studies have claimed evidence for a dip in the PPS at $k < 0.001$ Mpc$^{-1}$,
but our parametrization does not probe this range because these wave vectors correspond to such large
scales that the data are expected to have very little constraining power.
We do find a dip in the primordial power spectrum around $k = 0.002$ Mpc$^{-1}$.
This is driven by the deficit in the CMB temperature power spectrum, relative to the best-fit power-law model, around $\ell \approx 20$.

\begin{figure*}[htbp!]
\includegraphics[width=0.95\textwidth]{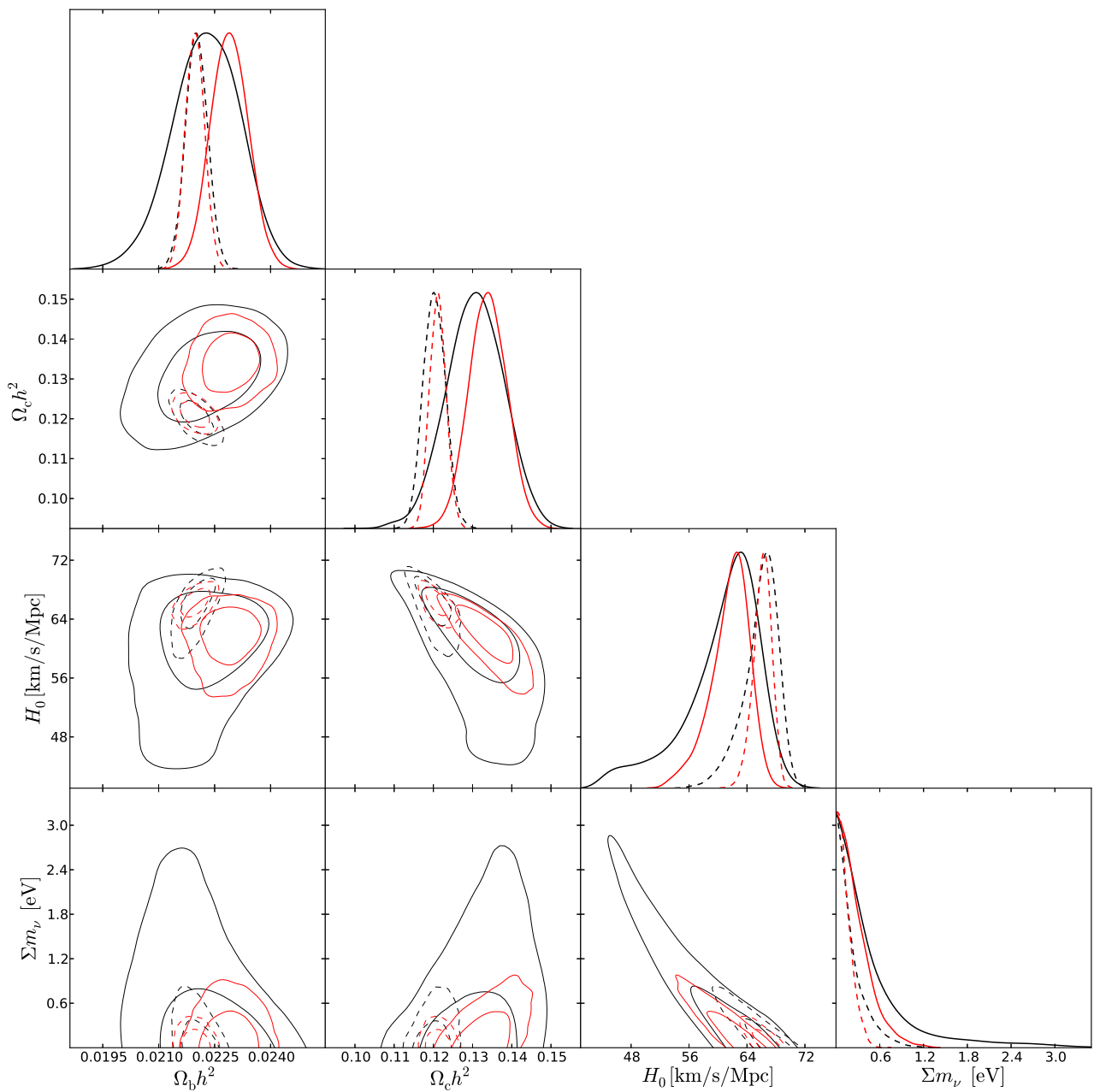}
\caption{
The posterior probability distributions of the cosmological parameters,
including neutrino mass. Results for the CMB-only data combination
are shown in black and those for CMB+BOSS in light red. 
The solid curves give the results with a free (splined) PPS,
while the dashed curves indicate results for the power law case.
The number of neutrino species is here fixed to the standard three. 
}
\label{fig:cosmo constraints cmb}
\end{figure*}

Figure~\ref{fig:cosmo constraints cmb} displays the cosmological parameter constraints
both for the free PPS case (for which we discussed the PPS constraints themselves above) using solid black curves/contours,
and for a power law PPS using dashed black curves/contours (we will discuss the results shown in red in the next subsection).
The contours show the $68 \%$ and $95 \%$ confidence level (CL)
regions, while the one-dimensional distributions are the marginalized posterior probability distributions.
Allowing more freedom in the PPS relative to a power law spectrum causes both a shift in
the best-fit/mean parameter values as well as a widening of the distributions.

\begin{figure*}[htbp!]
\includegraphics[width=0.48\textwidth]{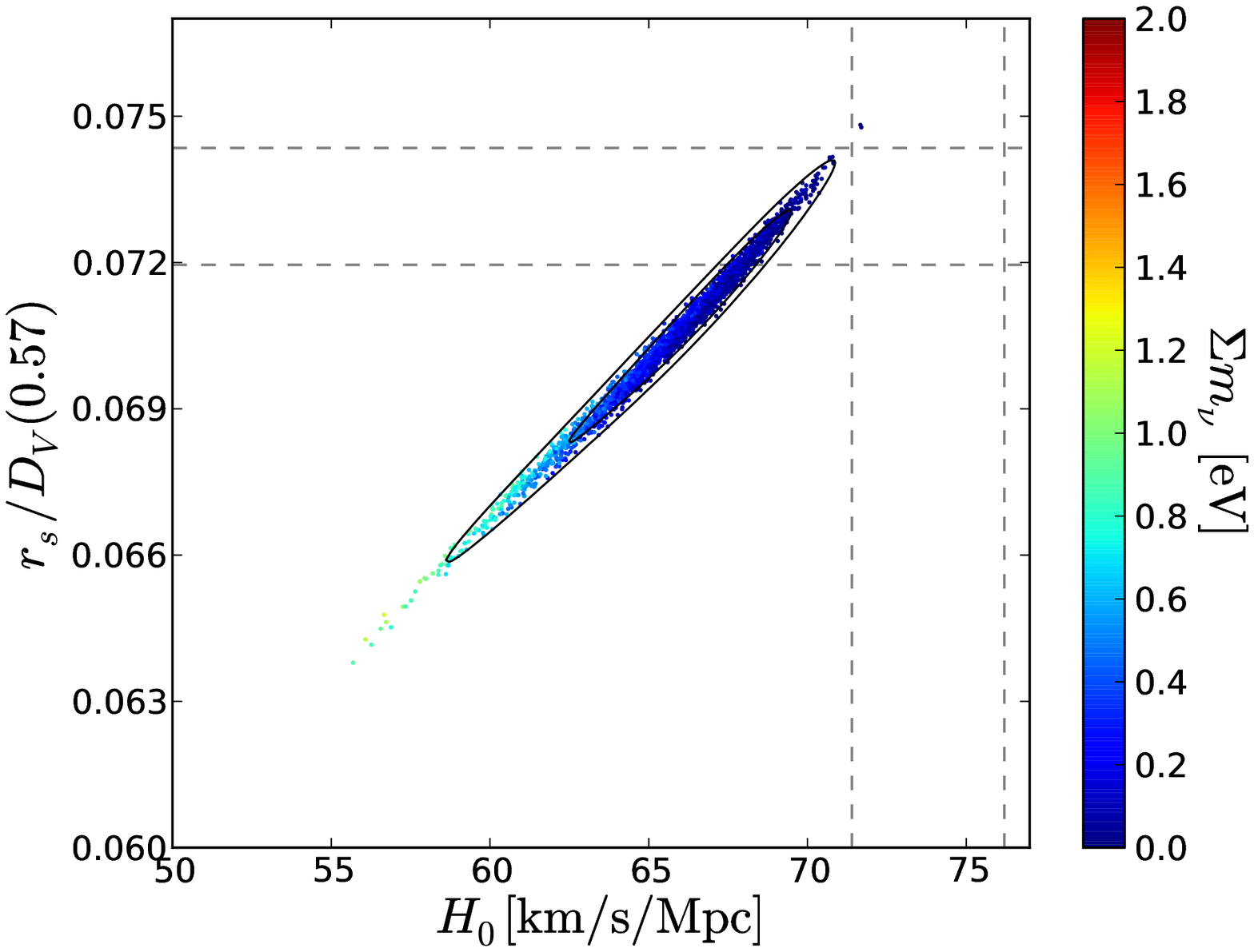}
\includegraphics[width=0.48\textwidth]{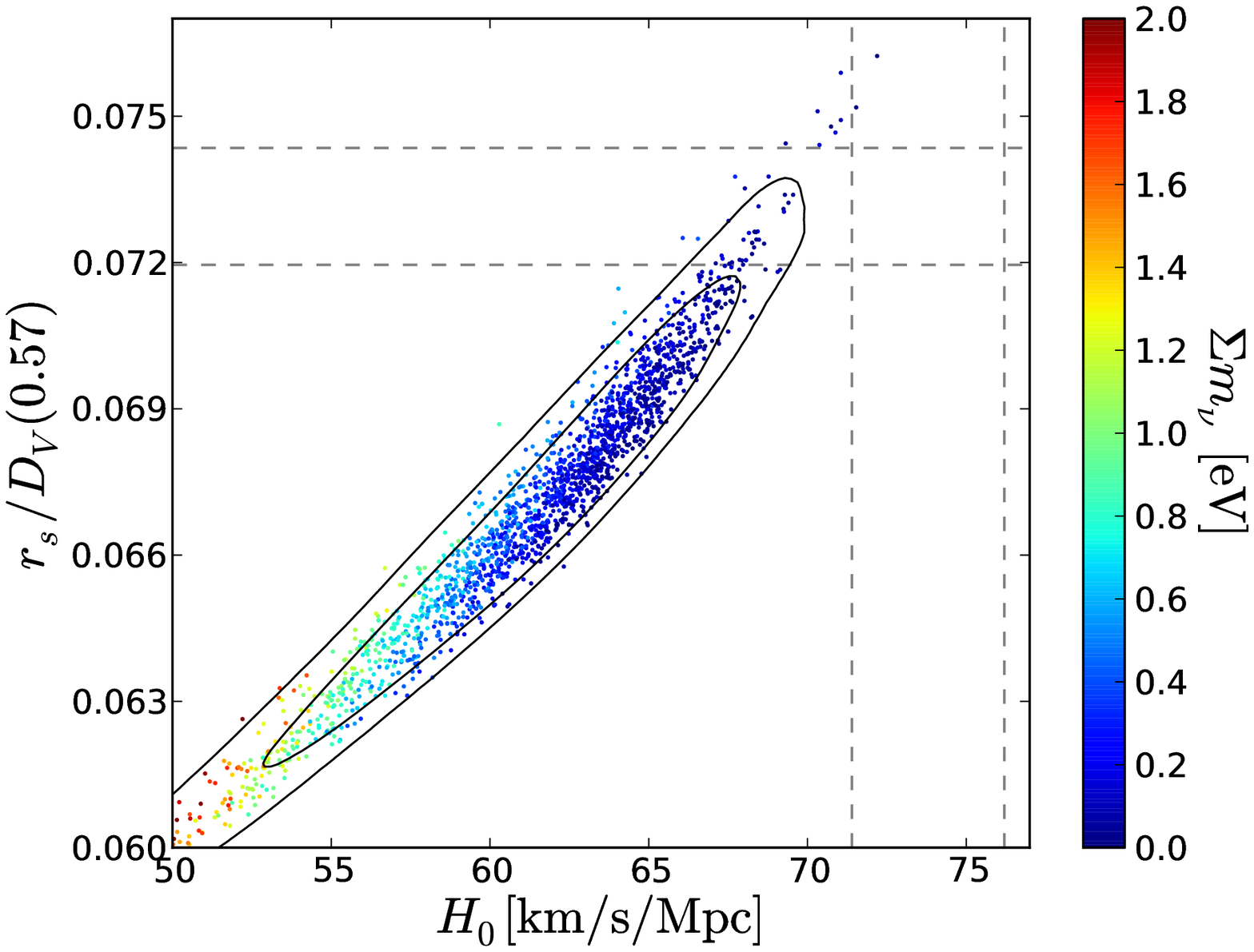}
\caption{Posterior distribution, for the CMB-only data combination, of
late-universe ``observables''
$H_0$ and $r_s/D_V(z=0.57)$, with $\mnu$ color coded.
Here $r_s$ is the sound horizon scale and $D_V(z=0.57)$
an effective distance to $z=0.57$, as measured from the angle-averaged BAO scale in the BOSS CMASS sample. The contours indicate $68 \%$
and $95 \%$ confidence regions and the dashed lines indicate $1\sigma$ ranges from direct measurement of $H_0$ and $r_s/D_V$ (see text).
Note that our analysis uses the full shape of the galaxy power spectrum rather than just the BAO measurement.
{\it Left:} Power law PPS. {\it Right:} Splined PPS. Allowing more freedom in the PPS causes a broadening of the distribution,
but retains the strong parameter (anti-)correlations so that the addition of a galaxy clustering and/or $H_0$ measurement will still tighten the neutrino mass constraint.}
\label{fig:h0-bao-mnu}
\end{figure*}

For the neutrino mass, the data are consistent with $\sumu = 0$ for both choices of the PPS,
but a free (splined) PPS significantly loosens the upper bound from
\be
\mnu < 0.63 \, {\rm eV} \quad {\rm (power \, \, law)} \nonumber
\ee
to
\be
\mnu < 1.9 \, {\rm eV} \quad {\rm (spline)} \nonumber
\ee
both at $95 \%$ CL.
This is as expected from Figure \ref{fig:pps cl mnu}, where the free PPS 
could compensate for neutrino mass effects while the restricted power law 
case strongly disfavored larger neutrino masses (despite the ability to 
adjust non-inflationary cosmological parameters). 

Let us discuss the constraints in more detail below.
In the following, we will often distinguish between ``late-universe'' parameters on the one hand, $H_0$ and $\mnu$,
and the parameters traditionally constrained very tightly by the acoustic peak structure of the CMB, $\omega_b$ and $\omega_c$,
on the other hand.
It is instructive to take the power law case (dashed black curves in Figure \ref{fig:cosmo constraints cmb})
as our starting point. In this case, the PPS is essentially featureless,
so that features in the observed CMB spectra directly tell us about the
transfer functions, which encode the rich physics of acoustic oscillations
during the drag epoch and of the growth of structure at late times,
and thus carry a wealth of information on cosmological parameters.
Indeed, the CMB peak structure allows for precise measurements of the baryon and cold dark matter
densities $\omega_b$ and $\omega_c$, as shown in Figure \ref{fig:cosmo constraints cmb}.

Late-universe ($z < 1100$) physics is mainly constrained by the distance to the CMB last scattering surface
and by CMB lensing, and to a lesser extent by the Integrated Sachs-Wolfe (ISW) effect. 
Still, in the presence of massive neutrinos, significant freedom remains in the parameter direction
corresponding to simultaneously varying $\mnu$ and $H_0$. This degeneracy can be understood
in the following simplified picture (see, e.g., \cite{houetal12,planckcosmoparam}).
Assume that the CMB measures $\omega_b$ and $\omega_c$ well, independently of neutrino mass.
Then, if we increase $\mnu$, the main effect on the CMB, {\it ceteris paribus}, is a decrease in the distance to the last
scattering surface of the CMB\footnote{A shift in the distance to CMB last scattering quickly worsens the fit to the data, as the angular size of the sound horizon,
which is the ratio of the sound horizon scale and
the distance to last scattering, $r_s/D_{LSS}$,
is measured very accurately. Note the effect of neutrino mass on $r_s$ is small (in the power law PPS scenario,
we are in the regime where neutrinos become non-relativistic {\it after} CMB last scattering).}
because neutrino mass increases the neutrino energy density and thus the expansion rate.
However, this shift in distance can be compensated by simultaneously lowering the dark energy density $\Omega_\Lambda$,
and thus $H_0$. This explains the degeneracy direction
seen in the $H_0 - \mnu$ panel of Figure \ref{fig:cosmo constraints cmb}. When the distance to last scattering is kept
constant like this, the joint variation in $\mnu$ and $H_0$ has a remaining effect on the ISW signal \cite{houetal12} and
CMB lensing, so that the $\mnu-H_0$ degeneracy is not exact and the CMB can still place a meaningful upper limit on
neutrino mass. The above explanation is of course only approximate, as in reality there is also some degeneracy with $\omega_b$
and $\omega_c$.

The late-universe degeneracy of $\mnu$ vs.~$H_0$ (or other combinations of late-universe parameters)
will play an important role in understanding how the neutrino and other
constraints improve when low-redshift data (galaxy clustering and/or a direct $H_0$ measurement) are added.
We therefore illustrate this CMB-only degeneracy in Figure \ref{fig:h0-bao-mnu}. The left panel depicts the power law PPS case.
It shows a scatter plot
of $H_0$ vs.~the ratio $r_s/D_V(z=0.57)$, where $r_s$ is the sound horizon scale, and $D_V(z=0.57)$ the volume-weighted distance to
$z=0.57$. For fixed $\omega_b$ and $\omega_c$, the latter quantity can be derived from
$\mnu$ and $H_0$ (i.e.~it is not an independent parameter). The ratio $r_s/D_V(z=0.57)$ will be useful for analysing the tightening of constraints
when including galaxy clustering
in the BOSS CMASS sample, where much of the information comes from the BAO measurement of $r_s/D_V(z=0.57)$.
The colors indicate the values of $\mnu$. The correlation between the three quantities plotted is clear from the figure:
increasing $\mnu$ leads to decreasing $H_0$ and $r_s/D_V(z=0.57)$. The horizontal and vertical bands indicate the $1\sigma$ ranges
for the measured BAO scale from the BOSS CMASS sample \cite{andersonetal12} and for $H_0$ \cite{hst11}. We will discuss
in the following sections how the BAO measurement relates to
the CMASS data used in this work, and how the CMASS data and $H_0$ prior help to constrain $\mnu$.

Let us now consider what happens when we allow for a splined PPS (right 
panel of Fig.~\ref{fig:h0-bao-mnu}, and solid black in 
Fig.~\ref{fig:cosmo constraints cmb}). 
The features now allowed in the PPS are partially degenerate with the effect of the transfer functions,
thus affecting the cosmological parameter constraints.
Specifically, we see that $\omega_b$ and $\omega_c$ become significantly less tightly constrained and that the mean
of (especially) $\omega_c$ is shifted to a larger value. Note however that these shifts, and those in the other parameters,
are consistent, in the sense that the mean value in the power law scenario is always within the range of values
allowed by the splined PPS posterior. This makes sense as the power law PPS is effectively embedded within the spline parametrization
(one can adjust spline parameters to very closely approximate any power law spectrum).
Since the power law case gives a good fit to the data, its mean cosmological parameter values should still give a good fit
in the splined PPS scenario.

Moving on to the late-universe parameters, we see that the mean of $H_0$ shifts down (in accordance with the anti-correlation
with $\omega_c$) and that $H_0$ obtains a larger uncertainty.
Finally, the additional PPS freedom  allows for larger values of $\mnu$. 
Given the anti-correlation with $H_0$ discussed above, this is what would be expected based on both the broadening
of the $H_0$ posterior and the downward shift of the mean of $H_0$.
The degeneracy directions of $\mnu$ with the other cosmological parameters remain
approximately the same when going from a power law to a splined PPS,
and the contours are simply widened in all directions (and shifted). For example, as $\mnu$ goes up,
$H_0$ has to decrease to maintain the best possible fit to the CMB data, as in the power law case.
Thus, the same physics (e.g.~keeping the distance to last scattering constant) still explains
the interplay between parameters, but the free PPS allows for larger parameter variations in all directions.
This is also shown in the right panel of Figure \ref{fig:h0-bao-mnu}, which presents
the joint constraints on $\mnu$, $H_0$
and $r_s/D_V(z=0.57)$ for the case of a free PPS. This figure will be useful when considering the effect
of adding the BOSS and/or $H_0$ data.

To summarize the main result of this subsection, the neutrino mass bound from CMB-only data strongly depends
on assumptions made about the PPS. If the PPS is restricted to a power law 
form, a strong upper bound on $\mnu$ is obtained. However, taking an agnostic 
approach with regards to the inflationary specifics of the primordial 
density fluctuations by allowing for a free form PPS largely undoes the
ability of CMB data to provide detailed information on neutrino mass.

\subsection{CMB + BOSS galaxy power spectrum constraints on neutrino mass}

\begin{figure}[htbp!] 
\includegraphics[width=\columnwidth]{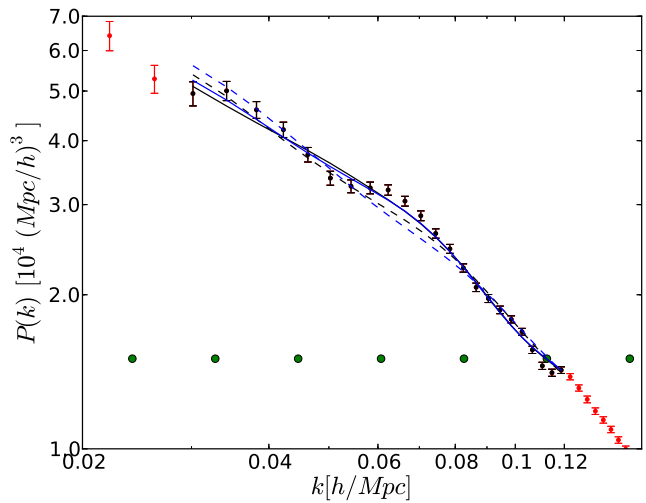}
\caption{The measured galaxy power spectrum of the BOSS CMASS sample is 
plotted as dots with error bars. 
The black data points are the only ones used in our analysis, spanning the range 
$k = 0.03 - 0.12\,h\mpci$.
The solid curves (black for spline PPS, blue for power law PPS) 
represent the models that best fit the {\it CMB}-only data
for fixed $\mnu=0$, while the dashed curves give the predictions by the models
best fitting the {\it CMB} data for $\mnu = 2.5$ eV (same as in 
Figure \ref{fig:pps cl mnu}). 
The green dots again indicate the locations of the PPS spline nodes.
}
\label{fig:pps pkg}
\end{figure}

As mentioned previously, however, the PPS enters differently into the 
matter density power spectrum, so we now investigate inclusion of the 
galaxy power spectrum of the BOSS CMASS sample in the data used. 
The measurements are shown in Figure \ref{fig:pps pkg}, with the 
range used for our analysis colored black. 
The green dots again indicate the PPS node wave vectors\footnote{Note 
that we show the power spectrum obtained from the data
assuming a fiducial cosmology; the inferred spectrum in 
a different cosmology is a shifted version
of the plotted spectrum (both in the horizontal and vertical directions).
The mapping between the wave vector in the plot and true wave vector (which
is used in the PPS parametrization) is thus cosmology dependent. To allow a direct, but cosmology dependent, comparison,
we have converted the node wave vectors to units $h/$Mpc, using the value of $h$ in the fiducial cosmology
of the BOSS analysis.}.

For illustration, the solid lines in Figure \ref{fig:pps pkg}
show the best-fit spectra to the CMB-only data set for fixed
$\mnu = 0$, i.e.~for the same cosmologies as the solid lines in Figure
\ref{fig:pps cl mnu}. The dashed lines show the best-fit spectra to the CMB-only data
for fixed $\mnu = 2.5$ eV (also as in Figure \ref{fig:pps cl mnu}). While the $\mnu = 0$
spectra for the two PPS treatments provide a decent fit to the galaxy clustering
data and are very similar, increasing neutrino mass to $\mnu = 2.5$ eV
significantly worsens the fit in both cases, although less in the case of a free PPS.
This suggests that, with the large scale structure data included,
neutrino mass can be constrained meaningfully
even in the free (splined) PPS case (although not as well as in the power law PPS scenario); 
as we have seen in the previous section that is not the case with {\it CMB} data only.
We will see below that this improvement indeed holds. 

The constraints from {\it CMB + BOSS\/} on the spline PPS
are shown in red in Figure \ref{fig:pps constraints p(k)}.
For most nodes, both the mean values and uncertainties
are very similar to those from {\it CMB\/}-only,
with only slight improvements in the uncertainties.
Thus, the current galaxy clustering data do not have a strong effect on 
primordial power spectrum constraints. 

However, the galaxy power spectrum data do have a strong impact on the 
cosmological parameter constraints. 
The posterior distributions from the {\it CMB + BOSS\/} analysis 
are shown in Figure~\ref{fig:cosmo constraints cmb} in red. 
In the case of a power law primordial spectrum (dashed), the neutrino mass bound is 
now
\be
\mnu < 0.34 \, {\rm eV} \quad {\rm (power \, \, law)} \nonumber
\ee
at 95\% CL, an improvement by almost a factor 2 from the
CMB-only case.
Note that our bound is slightly stronger than the result 
$\mnu < 0.39\,$eV from combining CMB + BOSS CMASS in \cite{giusarmaetal13}; 
the difference is due to their inclusion of the reconstructed CMB lensing 
power spectrum from Planck \cite{plancklens13}, which prefers a larger $\mnu$ and 
thus weakens the upper limit somewhat (see also \cite{planckcosmoparam}). 

\begin{figure*}[htbp!]
\includegraphics[width=0.95\textwidth]{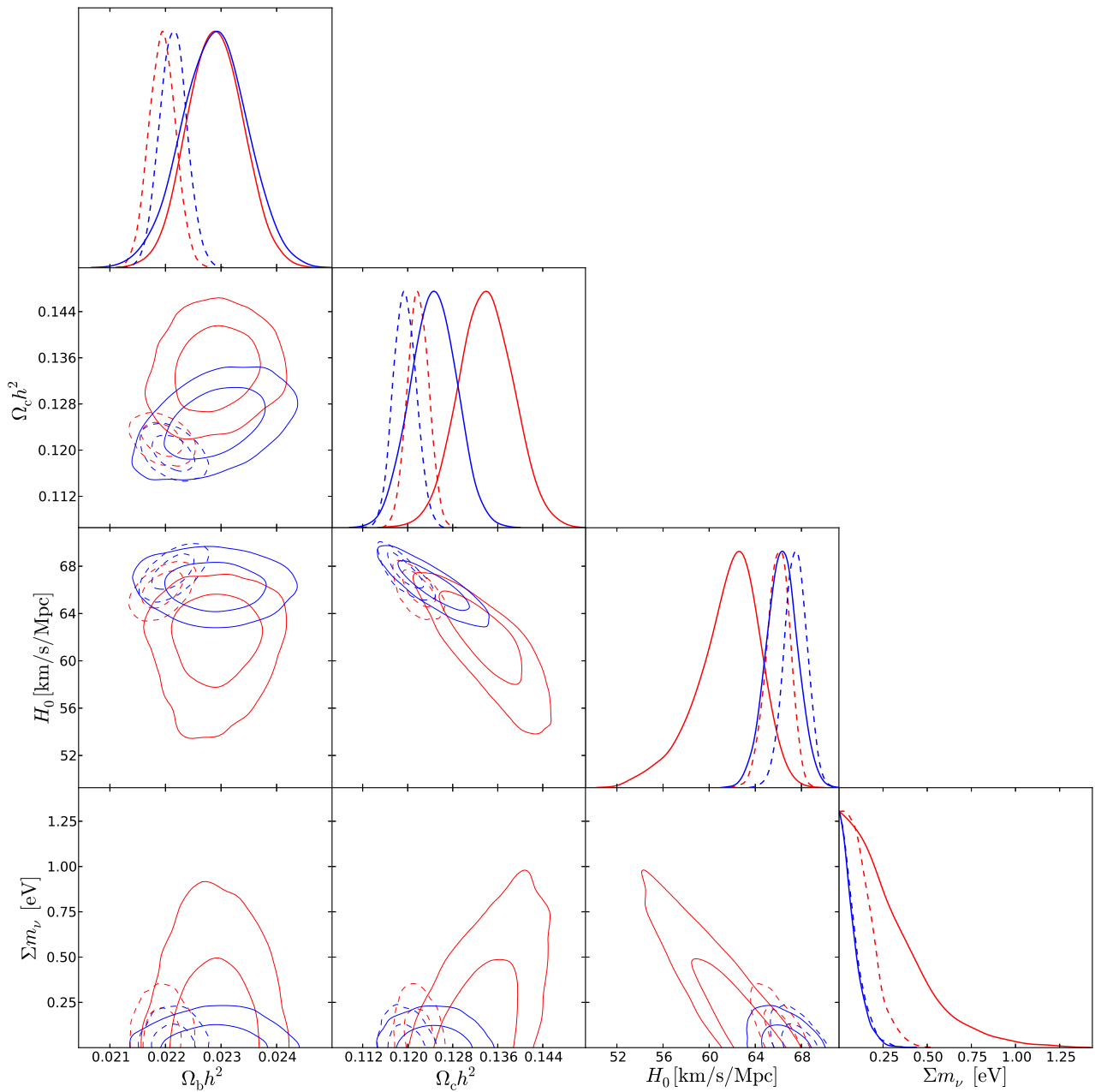}
\caption{
The posterior probability distributions of the cosmological parameters,
including neutrino mass, when CMB data are combined with low-redshift observations.
Note the change of scale from Figure \ref{fig:cosmo constraints cmb}. 
Results
for the {\it CMB + BOSS\/} data combination are shown in light red, while
the {\it CMB + BOSS + $H_0$\/} case is shown in dark blue.
In both cases, solid curves depict the free (splined) PPS and dashed the power law PPS.
The number of neutrino species is here fixed to the standard three.
}
\label{fig:cosmo constraints cmb+boss}
\end{figure*}

For the free PPS case, we find that the neutrino mass bound from 
{\it CMB + BOSS\/} data becomes
\be
\mnu < 0.72 \, {\rm eV} \quad {\rm (spline)} \nonumber 
\ee
at 95\% CL, a 
factor 2.6 stronger than the CMB-only case.
Thus, the galaxy power spectrum is able to rein in the effect of PPS freedom. 
Indeed this bound is comparable to having no 
galaxy data but restricting to a power law PPS.

To gain insight into the tightening of the neutrino mass bound,
let us first consider the power law PPS case, which is relatively easy to understand.
In this case, to a good approximation, the information in the BOSS power spectrum
can be represented by a measurement of the BAO scale,
given for the DR9 CMASS sample in \cite{andersonetal12} by $D_V(z=0.57)/r_s = 13.67 \pm 0.22$.
While this measurement ignores any information on the scale dependent suppression
of power due to neutrinos, and is based on a larger range of scales than used in our analysis,
we have checked explicitly that the constraints on all parameters from this BAO measurement
(combined with CMB) are very similar to those using
the full galaxy power spectrum (see also, e.g., \cite{giusarmaetal13}). 

Since $\omega_b$ and $\omega_c$ are already well measured from CMB-only,
and are very weakly affected by the BOSS measurement, we focus on the late-universe parameters
in Figure \ref{fig:h0-bao-mnu} (left panel). The $1\sigma$ allowed range for $r_s/D_V(z=0.57)$ (from now on just $r_s/D_V$)
from \cite{andersonetal12} is shown by the horizontal bands.
We have already discussed the CMB-only $\mnu - H_0$ degeneracy in the previous
subsection, and the anti-correlation of $\mnu$ with $r_s/D_V$ can be understood in the same way.
An increase in $\mnu$ needs to be compensated by a decrease in the dark energy density to keep the distance to CMB last scattering fixed,
which in turn leads
to a longer distance to $z=0.57$. Since the sound horizon $r_s$ is hardly affected when $\mnu$ is varied,
this leads to a smaller value of the ratio $r_s/D_V$.
It is now obvious from the degeneracy direction shown in the left panel of Figure \ref{fig:h0-bao-mnu} and explained above,
why the BAO prior leads to the tightening of the $H_0$ posterior,
shift upward of its mean, and the strong improvement of the upper bound on $\mnu$ seen in
Figure~\ref{fig:cosmo constraints cmb}.
The smaller than expected shift in $H_0$
is due to the difference between the BAO prior
and the actual galaxy clustering measurement used in the chains. Indeed, when we replace the galaxy power
spectrum measurement with the BAO measurement (not shown), the $H_0$ posterior shifts to slightly larger
values.

When the PPS is parametrized by a spline (solid curves in Figure \ref{fig:cosmo constraints cmb}),
the CMB-only (black) constraints are weaker than in the power law scenario
and the inclusion of galaxy clustering data (red) tightens even the $\omega_b$ and $\omega_c$ posteriors.
Regarding the late-universe parameters, the right panel of Figure \ref{fig:h0-bao-mnu} shows
a similar situation to the power law case (left panel) for the CMB-only data combination:
while the allowed ranges of $\mnu$, $H_0$ and $r_s/D_V$
are significantly widened, there is a strong anti-correlation between neutrino mass and $r_s/D_V$ (and $H_0$),
due to the need for the dark energy density to compensate for the effect of $\mnu$ on the distance to last scattering.
Treating the galaxy power spectrum measurement as a BAO prior,
as we did for the power law PPS,
would thus again explain the strong improvement
in the neutrino mass bound seen in Figure \ref{fig:cosmo constraints cmb} and quoted above.
In other words, the larger range of allowed $\mnu$ values in the splined PPS case,
as compared to the power law case,
is in large part caused by extending the degeneracy direction with $H_0$ and $r_s/D_V$ (moving further along the diagonal to
the bottom left in Figure \ref{fig:h0-bao-mnu}),
and can thus largely be undone by a prior on $r_s/D_V$ (or $H_0$).

An important caveat is that, in the splined PPS case, replacing the galaxy power spectrum measurement with a BAO prior
is less justified than in the power law case because, in principle, freedom in the PPS can be (ab)used
to mimic or shift the acoustic peak in the galaxy power spectrum, which could result in a completely
wrong estimate of $r_s/D_V$.
In practice, the inclusion of CMB data significantly restricts the allowed variations in the PPS
so that information on $r_s/D_V$ is encoded in the galaxy
clustering data even with a splined PPS.
Indeed, we find that the error bar on $r_s/D_V$ improves by a factor two
relative to the CMB-only case when
the BOSS data are added (from $r_s/D_v=0.0662 \pm 0.0037$ to $0.0667 \pm 0.0018$). On the other hand, the resulting uncertainty is still about $50 \%$
larger, and the best-fit value significantly smaller, than the direct measurement from \cite{andersonetal12} ($r_s/D_V = 0.0732 \pm 0.0012$).

Based on the above arguments, and since an exact description of the parameter (and PPS) direction(s) constrained by the galaxy power spectrum
would be very complex and most likely not that helpful, we simply note that a description
in terms of a prior on $r_s/D_V$ is an insightful approximation
and does {\it qualitatively\/} reproduce the effect
of the BOSS power spectrum data on the parameter constraints. As in the case of the power law scenario,
the BOSS data induce a tightening of the $H_0$ and $\mnu$ bounds and shift the mean values of $H_0$
(slightly) up and the mean value of $\mnu$ down,
as expected from the degeneracy directions depicted in the right panel of
Figure \ref{fig:h0-bao-mnu}.
For comparison, we have also calculated the posteriors that would be obtained if the direct BAO measurement from \cite{andersonetal12} could be used in the splined PPS case
(replacing the galaxy power spectrum
measurement) and found that the parameter constraints using the BAO prior
would be significantly stronger than the true constraints from the galaxy power spectrum (e.g.~$\mnu < 0.32$ eV instead of $\mnu < 0.72$ eV),
confirming that much of the BAO information in the galaxy power spectrum gets lost due to the additional freedom in the PPS and that, unlike in the
power law PPS case, treating the galaxy power spectrum measurement as a measurement of the BAO scale is not a {\it quantitatively\/} accurate approximation.

In summary, combining measurements of cosmological perturbations at redshift 
$z \sim 1100$ and at low redshift ($z \sim 0.57$) provides valuable information 
on neutrino mass (and other parameters) even without assuming a form for the 
primordial power spectrum.
The influence of neutrino mass on expansion, rather than the free streaming suppression
of the matter power spectrum, is the dominant effect for current large scale structure data.

\subsection{CMB + BOSS + $H_0$ constraints on neutrino mass} \label{sec:cmbgalh0} 

In the previous sections, and in Figures \ref{fig:cosmo constraints cmb} and \ref{fig:h0-bao-mnu},
we have explained and shown that $\mnu$ is strongly anti-correlated with $H_0$.
Moreover, when including the BOSS data with the CMB data,
with a splined PPS the preferred value of the Hubble parameter,
$H_0 = 61.6 \pm 2.7$ km/s/Mpc (68\% CL),
is low compared to the value obtained with the standard power law PPS,
$H_0 = 66.1 \pm 1.3$ km/s/Mpc (which in turn is slightly lower than the value when $\mnu$ is fixed to $0.06$ eV),
and even lower compared to the direct HST measurement discussed in
Section~\ref{sec:datamethod}, $H_0 = 73.8 \pm 2.4$ km/s/Mpc.
This means that including the HST $H_0$ measurement in our combination
of data sets should strongly tighten the upper bound on $\mnu$,
and {\it especially} so for a splined PPS.

Thus, it is worth investigating constraints from the {\it CMB + BOSS + HST}
data set. We show the results with the blue curves and contours in Figure
\ref{fig:cosmo constraints cmb+boss} and repeat the {\it CMB + BOSS} results 
from Figure~\ref{fig:cosmo constraints cmb} in red for comparison. We find that,
especially in the splined PPS case, the bounds on $\Omega_c h^2$, 
$H_0$ and $\mnu$ are all strongly affected. All the changes can be easily understood in terms of the 
degeneracies between each parameter and $H_0$, as shown by the red contours.
The neutrino mass upper bound becomes much stronger.
In the power law case the upper limit 
becomes
\be
\mnu < 0.19 \, {\rm eV} \quad {\rm (power \, \, law)} \nonumber
\ee
while for the spline case it becomes
\be
\mnu < 0.18 \, {\rm eV} \quad {\rm (spline)}\ . \nonumber
\ee
It is interesting to note that the $H_0$ prior is so powerful that the freedom in the PPS in the case of a splined
primordial spectrum no longer weakens the neutrino bound. 

The prior on $H_0$ has little influence on the constraints on 
the splined PPS itself; those from the {\it CMB + BOSS + HST\/} 
data combination are similar to those without the HST prior, so we do not show 
them separately in Figure~\ref{fig:pps constraints p(k)}.


We discussed above that the {\it CMB + BOSS} data combination prefers
a much lower $H_0$ value than the R11 Hubble constant measurement,
and that the discrepancy is more severe in the case of a free PPS.
This tension between the data sets (in the context of $\Lambda$CDM with massive neutrinos)
is also reflected in the goodness of fit. When $H_0$ is added to the {\it CMB + BOSS}
data combination, the fit worsens by $\Delta \chi^2 = 10.5$ (using the splined PPS),
while $\Delta \chi^2 \approx 1$ is expected if all data are consistent with a single underlying model.
The tight bounds presented above are largely driven by this tension
because neutrino mass is anti-correlated with $H_0$.
Because of this importance of the large value of the $H_0$ prior,
and because, as discussed briefly in Section \ref{sec:datamethod},
the tension between R11 and Planck
might point to the presence of additional errors in the direct $H_0$ measurement not included in the published uncertainty,
we next study briefly how the $\mnu$ bound is affected if the $H_0$ measurement is modified.

Replacing the R11 measurement by the revised $H_0$
prior of \cite{E13} ($H_0 = 72.5 \pm 2.5$ km/s/Mpc),
we find the upper limit $\mnu < 0.21$ eV, both for a power law and for a splined
PPS. The constraint is thus only slightly weakened and it remains true that
the inclusion of the Hubble prior renders the neutrino mass limit insensitive to the choice
of PPS model. However, if we instead use the value given in \cite{E13} based on the maser distance anchor
only ($H_0 = 70.6 \pm 3.3$ km/s/Mpc), the neutrino bound weakens to $\mnu \lesssim 0.27$ eV for both choices of PPS.
Since the neutrino mass bound has a non-negligible dependence on which $H_0$ value is used,
it will be extremely valuable to reach a robust, consensus Hubble constant measurement in the near future.

\subsection{Summary of neutrino mass bounds} 

Table~\ref{tab:mnubound} summarizes the 95\% CL upper bounds obtained 
on $\mnu$ for the various combinations of data sets and PPS cases. 
We see that inflationary freedom strongly affects neutrino mass bounds. 
Constraining the PPS through multiple types of observations, such as 
the CMB temperature power spectrum and galaxy power spectrum together, helps 
considerably. Further adding an external constraint on the Hubble constant 
compensates almost totally for the added inflationary freedom, allowing 
a more inflationary model independent bound.

\begin{table}[!htb]
\begin{tabular}{l|ccc} 
PPS & CMB & CMB+BOSS & CMB+BOSS+$H_0$ \\ 
\hline 
Power law$\ $&\ $\mnu < 0.63$\ & $\ \mnu < 0.34\ $ & $\ \mnu < 0.19\ $ \\ 
Spline & $\mnu < 1.9$ & $\ \mnu < 0.72\ $ & $\ \mnu < 0.18\ $ \\ 
\end{tabular}
\caption{The 95\% CL upper bounds on $\mnu$, in eV, are listed for the various 
combinations of data and theory models. The number of neutrino species is 
fixed at three.
} 
\label{tab:mnubound}
\end{table}

\subsection{Joint constraints on neutrino mass and number of species}
\label{subsec:mnu neff}

\begin{figure*}[htbp!]
\includegraphics[width=0.95\textwidth]{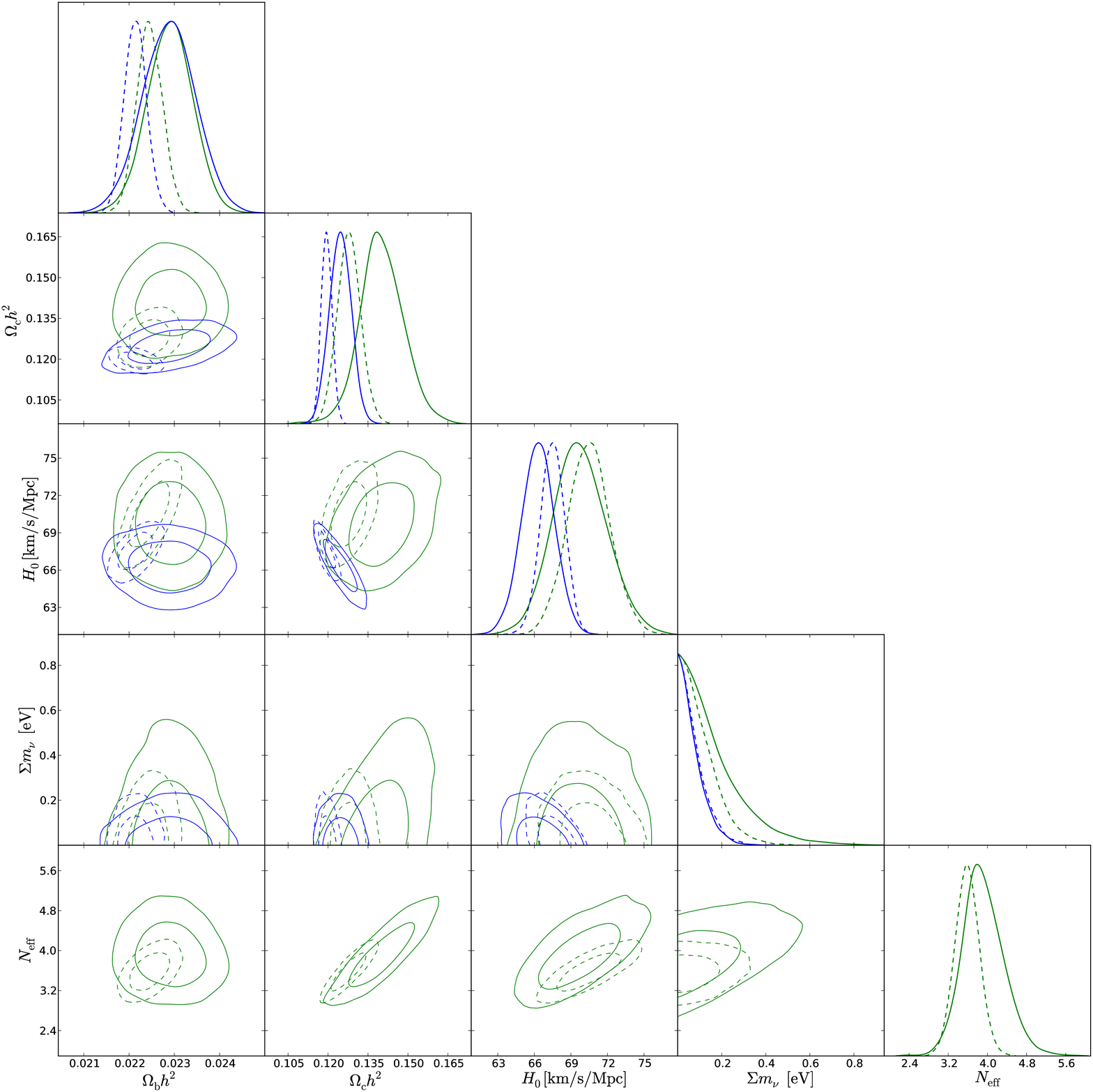}
\caption{ 
The posterior probability distributions of the cosmological parameters 
for the CMB+BOSS+$H_0$ data combination. The light green curves and contours present the
case with both the total neutrino mass 
$\mnu$ and the effective number of neutrino species $\neff$ allowed 
to vary, while the results in dark blue are for the fixed $\neff = 3.046$,
also shown in Figure \ref{fig:cosmo constraints cmb+boss}.
Note the change of scale from the previous figures.
In both cases, solid curves depict the free (splined) PPS and dashed the power law PPS. 
}
\label{fig:mnu neff all}
\end{figure*}

An important property of neutrinos in addition to their total mass is the 
effective number of neutrino species (quantifying the energy density of relativistic neutrinos
in terms of the energy density of a neutrino that has decoupled completely before electron-positron
annihilation). In the standard picture this is $\neff=3.046$. This can be altered either by adding more 
species, e.g.\ sterile neutrinos, or adding more energy density, e.g.\ by 
changing the neutrino thermal history or even having other contributions to the (free-streaming)
relativistic energy density (in which case the more general term {\it dark radiation} applies).
Sterile neutrinos in particular have received a lot of recent attention (see \cite{sterneutwhitepaper12}
and references therein), as the addition of one or two light sterile neutrinos may help explain observed anomalies in
short baseline neutrino oscillation experiments. Moreover, $\neff > 3.046$ would ameliorate
the tension discussed briefly above between the preferred value of $H_0$ from CMB data and direct measurements of $H_0$
(see also \cite{planckcosmoparam,wymanetal13,battyemoss13,haha13,gagiula13}).

We therefore now include $\neff$ as a free parameter and examine the 
constraints on its value\footnote{Big bang nucleosynthesis also constrains $N_{\rm eff}$,
see e.g.~\cite{nollettsteig13}, but we do not include these data in this study.}, as well as the effect on the neutrino mass bound. 
When $\neff$ is a free parameter {\it and} the PPS is described by a spline, the freedom in parameter space is so large
that we can only obtain robust constraints when all data are combined. We thus only show results
for the {\it CMB + BOSS + $H_0$} data combination.
The green solid (splined PPS) and dashed (power law PPS) curves and contours in Figure \ref{fig:mnu neff all} show
the posterior distributions with free $\neff$.
For comparison, we show in blue the results for the same data combination when $\neff$ is fixed to the standard value. 

Table~\ref{tab:mnuneff} summarizes the constraints. Looking first at the mean and standard deviation
of $\neff$ (second column), we find that the {\it CMB + BOSS + $H_0$} data have a mild preference for $\neff$
larger than the standard value, at slightly more than $95 \%$ CL significance.
This is largely
driven by the large value of the direct measurement\footnote{For example, in the power law case, using only CMB
data or CMB with a BAO measurement, the $\neff$
measurement is consistent with $\neff = 3.046$ at the $95 \%$ CL \cite{planckcosmoparam}.} of $H_0$
in combination with the strong correlation between $H_0$ and $\neff$.
The splined PPS case prefers a slightly larger $\neff$ and constrains its value less tightly than the power law PPS case.
Also the upper bound on the neutrino mass is weaker for the spline PPS than for a power law,
while both are weaker than the bounds obtained for fixed $\neff = 3.046$ (shown in parentheses).
Thus, unlike in the case of fixed $\neff$, when $\neff$ is a free parameter, even the {\it CMB + BOSS + $H_0$} constraints are weakened by
allowing additional freedom in the PPS.

The physics behind the $\neff$ constraint can be understood in the usual way
when the PPS follows a power law (e.g.~\cite{houetal12,planckcosmoparam}).
In order to fit the CMB temperature power spectrum, an increase in $\neff$ needs to be accompanied by an
increase in $\omega_c$ to keep the matter-radiation equality scale constant, and by an increase in $H_0$
to keep the angular size of the sound horizon constant (since the increase in $\neff$ decreases $r_s$).
Moving along this degeneracy direction in parameter space, the dominant remaining effect on the CMB is that the
angular size of the Silk damping scale decreases (leading to more damping), making it possible even for CMB-only data
to constrain $\neff$.
The strong correlation between $\neff$ and $H_0$ discussed above explains how adding an $H_0$ (and galaxy clustering) measurement
to the CMB data strongly tightens the $\neff$ bound.
The above explanation mostly also applies to the splined PPS case, leading to the same approximate parameter degeneracy directions
(except for $\omega_b$), while the extra freedom in the PPS simply broadens (and slightly shifts) the contours.
Finally, we note that $\neff$ and $\mnu$ are only weakly correlated with each other.

\begin{table}[!htb]
\begin{tabular}{l|c}
PPS & CMB+BOSS+$H_0$ for $\neff$ free \\
\hline
Power law$\ $ & $\ \mnu < 0.26$, $\neff = 3.59 \pm 0.25\ $ \\
Spline & $\ \mnu < 0.43$, $\neff = 3.92 \pm 0.42\ $ \\ 
\end{tabular}
\caption{The 95\% CL upper bounds on $\mnu$, in eV, and the mean and standard 
deviation for $\neff$, fitting for both simultaneously, are listed for the 
CMB+BOSS+$H_0$ combination of data. Recall from Table~\ref{tab:mnubound} 
the corresponding neutrino mass constraints for fixed $\neff=3.046$ 
are 0.19 eV for power law and 0.18 eV for spline PPS. 
}
\label{tab:mnuneff}
\end{table}

\section{Summary \& Conclusions}
\label{sec:concl}

The universe on large scales provides a unique laboratory for studying fundamental properties of neutrinos.
While neutrino mass differences are well measured by more traditional particle physics experiments,
the most accurate bound on the absolute neutrino mass scale currently, and for the foreseeable future, comes from
cosmological data. It is therefore crucial to investigate to what extent
this measurement depends on the assumed cosmological model. 

One key ingredient of the assumed cosmology is the primordial power spectrum
of curvature perturbations. The strong bounds on neutrino mass quoted in the literature (e.g.~$\mnu < 0.23$ eV \cite{planckcosmoparam})
typically assume a power law PPS (sometimes with a running index). In this article, we have instead studied
cosmological neutrino constraints when no functional form is assumed for the PPS.
As a compromise between allowing as much freedom in the PPS as possible and computational practicality,
we have modeled the PPS by a spline with 20 free nodes (though the results 
are insensitive to the exact number).
We have derived constraints using a compilation of CMB data and have quantified the effect of including
low redshift measurements of the Hubble constant $H_0$ and galaxy clustering.

We found that CMB data alone constrains the PPS to better than $10 \%$ over a large range of wave vectors,
$k \approx 0.01 - 0.25$ Mpc$^{-1}$, as shown in Figure \ref{fig:pps constraints p(k)}.
No significant deviation from a power law is found.
The PPS constraint itself does not improve significantly when current 
low-redshift data are included.

The constraints on the sum of neutrino masses (Table \ref{tab:mnubound})
do depend strongly on whether or not low-redshift information is used.
For the CMB data set only, $\mnu$ is very poorly constrained when the PPS is left free,
giving a bound $\mnu < 1.9$ eV ($95 \%$ CL) compared to $\mnu < 0.63$ eV assuming a power law PPS.
However, when low-redshift data are added,
the neutrino mass bound becomes stronger and more robust against the choice of the PPS model.
Including the galaxy power spectrum from BOSS leads to a bound $\mnu < 0.72$ eV (splined PPS) compared to $\mnu < 0.34$ eV for a power law PPS.
Finally, when also a prior on $H_0$ from HST is incorporated, the mass limit becomes almost independent of the chosen
PPS model, and very strong: $\mnu \lesssim 0.18$ eV. 

We also derived joint constraints on neutrino mass {\it and} the effective number of neutrino species, $\neff$, which are
summarized in Table \ref{tab:mnuneff}. Combining all three probes,
we obtained strong bounds on both quantities, even with a free PPS.
Unlike in the case of fixed $\neff$, the extra freedom in the PPS does weaken the neutrino bounds relative to the
power law scenario, by approximately a factor of 1.65 on both the mass 
and number of species uncertainties. The data show a preference for $\neff$ larger than the
canonical value $\neff = 3.046$, but only at slightly more than $95 \%$ CL and strongly driven by the $H_0$
measurement.

In summary, we have found no strong deviations from a power law primordial power spectrum and have shown that, while with a free (splined)
PPS, CMB data alone hardly constrain $\mnu$, adding galaxy clustering or $H_0$ measurements enables strong neutrino
limits regardless of the primordial power spectrum model.


\acknowledgments

We thank Olga Mena for her assistance with the galaxy power spectrum
likelihood code and Jan Hamann for useful discussion regarding Appendix B.
Part of the research described in this paper was carried out at the 
Jet Propulsion Laboratory, California Institute of Technology, 
under a contract with the National Aeronautics and Space 
Administration. This work is supported by NASA ATP grant 11-ATP-090, 
DOE grant DE-SC-0007867 and the Director,
Office of Science, Office of High Energy Physics, of the U.S.\ Department
of Energy under Contract No.\ DE-AC02-05CH11231, and 
by Korea World Class University grant 
R32-2009-000-10130-0. 
RdP thanks the Institute for the Early Universe at Ewha University, Seoul,
where part of this work was performed,
for its hospitality.

\appendix

\section{Varying the PPS model}
\label{sec:nodes}

\begin{figure*}[htbp!]
\includegraphics[width=0.47\textwidth]{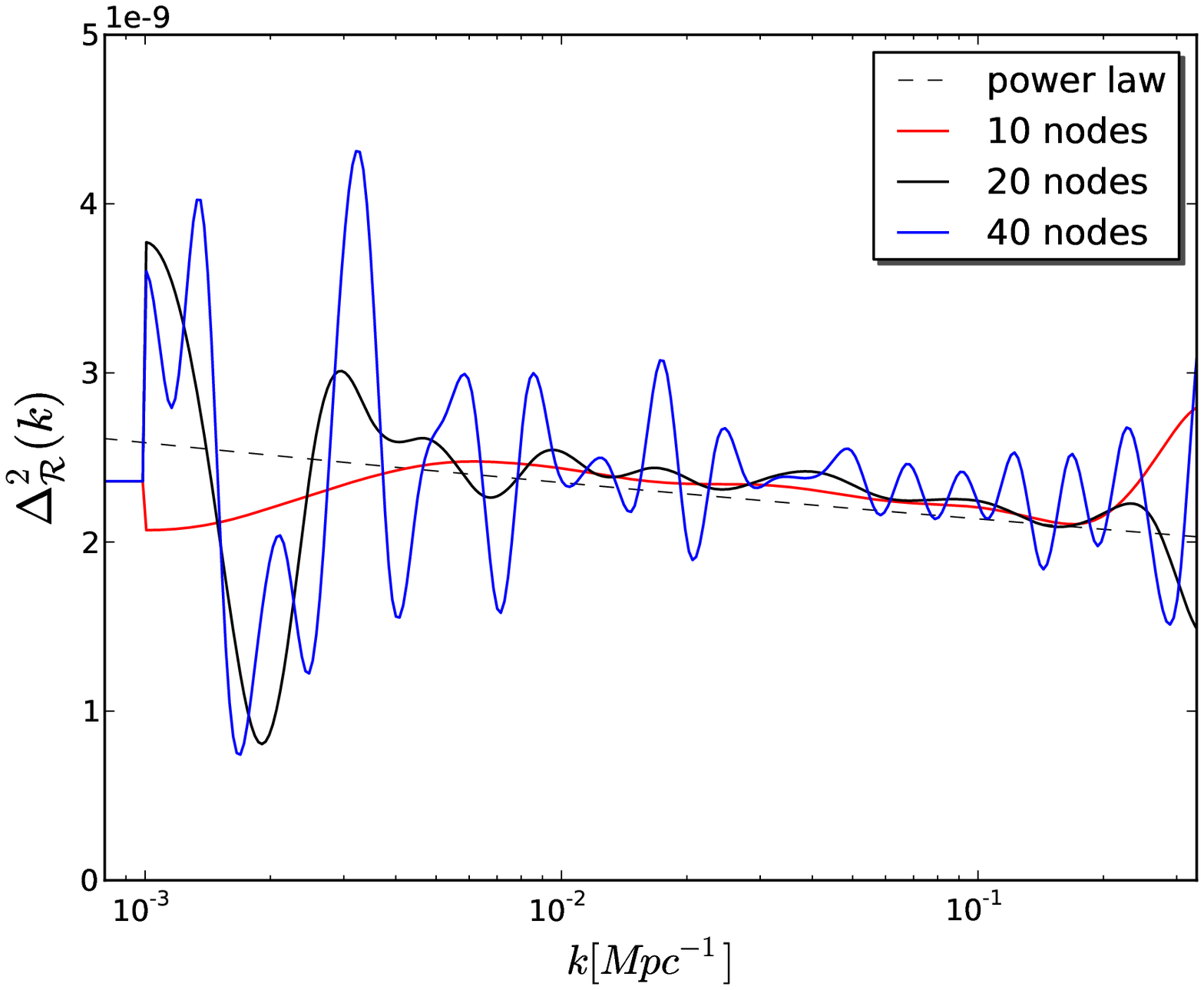}
\includegraphics[width=0.47\textwidth]{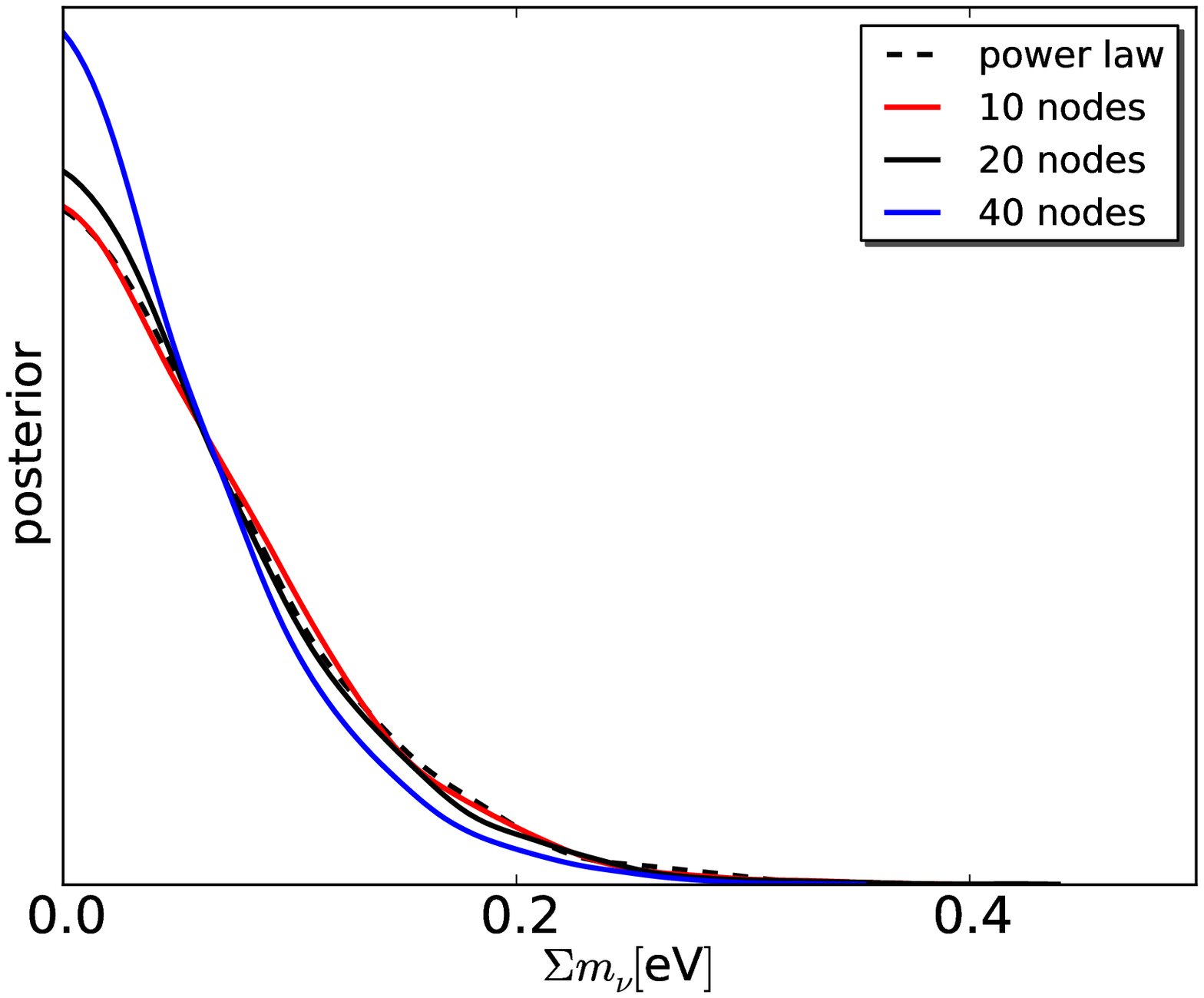}
\caption{{\it Left:} The mean posterior PPS for different choices of the number of spline nodes (with the best-fit power law PPS
shown for comparison), using the full {\it CMB + BOSS + $H_0$}
data combination. The number of neutrino species is fixed to the 
standard three.
Allowing more nodes makes the reconstructed individual PPS node values more noisy (error bars not shown). 
{\it Right:} The posterior distribution of the sum of neutrino masses for the cases shown in the left panel.
When all data are combined, the bound is remarkably robust against
varying assumptions about the PPS and to a first approximation stays constant.
The counterintuitive (but small) improvement in the $\mnu$ bound as freedom in the PPS is increased can be explained by
the fact that the CMB data prefer a lower $H_0$ when more freedom in the PPS is allowed, 
so that adding the higher $H_0$ (and galaxy clustering) measurement forces 
$\mnu$ along its degeneracy direction with $H_0$, to lower values.
} 
\label{fig:mnuvsnodes}
\end{figure*}

Our goal in this work has been to derive cosmological, and specifically neutrino, constraints when
no assumptions are made on the form of the primordial power spectrum. To this end,
we have modeled the PPS as a spline with 20 nodes logarithmically spaced in the range $k = 0.001 - 0.35$ Mpc$^{-1}$.
We chose this number of nodes because it leads to a large amount of freedom in the PPS,
allowing for features on approximately the same scales as those induced in the observed power spectra by the transfer functions for
matter and radiation perturbations. Moreover, this number of nodes is still small enough for the PPS to be well constrained
(to better than $10 \%$ for $k \approx 0.01 - 0.25$ Mpc$^{-1}$) and for it to be possible to obtain properly converged
MCMC results.

While our default parametrization is thus well motivated, it is interesting to see how the results change when the number of nodes
is varied. We have therefore also calculated constraints using 10 and 40 nodes (with the same $k$ range), using the full
{\it CMB + BOSS + $H_0$} data compilation.
The left panel of Figure \ref{fig:mnuvsnodes} shows the mean posterior PPS for these cases (and the best-fit power
law spectrum for comparison). As expected, the PPS choices with fewer nodes and hence less freedom approximately follow a smoothed
version of the ones with more nodes. While error bars are not explicitly shown to avoid clutter,
the uncertainties in the individual node values increase with increasing number of nodes.

The right panel of Figure \ref{fig:mnuvsnodes} shows the resulting posterior distribution of the sum of neutrino masses.
As already suggested by the good agreement between the power law and the 20-node spline neutrino limits, the $\mnu$ bound
is remarkably robust against changes in the assumed PPS model. We do note, however, that the posteriors of other parameters
undergo more significant shifts as the number of nodes is varied. Moreover, the robustness of the neutrino bound relies
on the use of low-redshift data to complement the CMB power spectra. The $\mnu$ limit depends more strongly on the PPS parametrization
when fewer data sets are used.

\section{The role of multiple transfer functions and of CMB polarization}
\label{sec:cmbpol}

We have seen in this article that access to multiple probes is crucial
for obtaining PPS-independent cosmology constraints.
When the combined data sets are measurements of cosmic perturbations, here in the
form of CMB and galaxy power spectra,
this can be understood qualitatively as follows (see, e.g., \cite{TegZal02}).
An observed power spectrum is the convolution of a transfer function with
the primordial power spectrum, with the relevant cosmological (e.g.~neutrino) information
encoded in the former.
If only one power spectrum is observed, the effects of the cosmological parameters
are in principle degenerate with variations in the PPS.
However, when multiple spectra, with differing transfer functions, are combined,
freedom in the PPS can in general not be used to undo the transfer functions effects
on all spectra simultaneously and PPS-independent transfer function information can be extracted.
As a simple example, if the matter power spectrum could be directly measured at two redshifts,
then the ratio of these power spectra would be explicitly independent of the PPS and would
give the transfer function of matter perturbations between the two redshifts, leading to constraints on
the dark energy density and neutrino mass.

The example of the complementarity described above that we have focused on in this article, is the combination
of the CMB power spectra with the galaxy power spectrum.
In this appendix, we note that even the CMB-only data set makes use of two types of perturbations, namely
temperature and E-mode polarization.
To see to what extent the inclusion of polarization data has provided PPS-independent cosmological information according to the above
description, we have run Monte Carlo chains with the WMAP polarization (WP) data set replaced by a prior on
the optical depth to reionization\footnote{In the case of a free PPS, we have also implemented a prior
$\tau = 0.097 \pm 0.015$, which is the free-PPS constraint on the optical depth with the WP data included.
This choice gives the same neutrino mass bound as the $\tau = 0.09 \pm 0.013$ prior.}, $\tau = 0.09 \pm 0.013$ (see also \cite{planckcosmoparam}).
In the power law PPS case, we find that the $\tau$ prior is a good approximation of the information carried by the WP data: 
the $\mnu$ bound only weakens slightly from $\mnu < 0.63$ eV to $\mnu < 0.83$ eV.
However, for the splined PPS, the neutrino bound weakens by a large amount when the WP data are replaced,
going from $\mnu < 1.9$ eV to $\mnu < 3.2$ eV.
Thus, without the E-mode polarization data, even when $\tau$ is still known as well as it would be {\it with} those data,
the CMB-only neutrino bound is extremely weak. The polarization data have therefore played a large role in our
CMB-only constraints for a free PPS. This is in agreement with our qualitative picture described above
of the importance of having access to multiple transfer functions, 
and bodes well for future data with full polarization information and 
measurements of galaxy clustering at multiple redshifts.

\bibliography{refs}

\end{document}